\documentclass[aps,letterpaper,twocolumn,pra,footinbib,floatfix,showpacs]{revtex4}
\usepackage{amsmath}
\usepackage{amsfonts}
\usepackage{amssymb}
\usepackage[scanall]{psfrag}
\usepackage{psfrag}
\usepackage{graphicx}
\usepackage{color}
\usepackage{subfigure}

\newcommand{\p}[1]{\left( #1 \right)}
\newcommand{\br}[1]{\left[ #1 \right]}

\renewcommand{\vec}[1]{\boldsymbol{#1}}
\newcommand{\dee}[2]{\frac{\partial #1}{\partial #2}}

\newcommand{\vk}{\vec{k}}

\newcommand{\dg}{^{\dagger}}

\newcommand{\gb}{\beta}
\newcommand{\gd}{\delta}

\newcommand{\go}{\omega}
\newcommand{\gO}{\Omega}

\newcommand{\mat}[1]{\begin{pmatrix} #1 \end{pmatrix}}
\newcommand{\grad}{\vec{\nabla}}

\begin{document}

\author{Yariv Yanay}
\affiliation{Laboratory of Atomic and Solid State Physics, Cornell University, Ithaca NY 14850}
\author{Erich Mueller}
\affiliation{Laboratory of Atomic and Solid State Physics, Cornell University, Ithaca NY 14850}
\title{Dispersion and wavefunction symmetry in cold atoms experiencing artificial gauge fields}
\date{\today}

\begin{abstract}
We analyze the single particle quantum mechanics of an atom whose dispersion is modified by spin orbit coupling to Raman lasers. We calculate how the novel dispersion leads to unusual single particle physics.  We focus on the symmetry of the ground state wavefunction in different potentials.
\end{abstract}


\maketitle

\section{Introduction}

One of the most exciting developments in cold atom experiments is the ability to emulate the Hamiltonians of charged particles in magnetic fields \cite{spielman1} and electrons with spin-orbit coupling \cite{spielman2}. These techniques allow one to tune the dispersion in complicated spatially dependent ways.  Here we show the that resulting single particle quantum mechanics is profoundly different than what we are used to.  For example, theorems about the number of nodes in the ground state \cite{qmtext} no longer apply, and by tuning  experimental parameters one can change the ground state in a double well from symmetric to antisymmetric, and back again.

In the experiments of Lin et al. \cite{spielman1,spielman2,spielman3,spielman4}, Rubidium atoms in the F=1 hyperfine manifold interact with two co-propagating lasers.  These lasers drive Raman transitions between the three magnetic hyperfine states $m=-1,0,1$.  Including the effect of the quadratic Zeeman field, the Hamiltonian in the rotating frame is
\begin{equation} \begin{split}\label{eq:threeD}
\hat H_{3} = &\p{\frac{\hbar^{2}}{2m}\hat{\vec k}^{2} + \frac{\gd}{2}}\mathbb I + \gd {\bf S}_{z} 
+ \hbar\omega_{q}\mat{ 1 & 0 & 0 \\ 0 & 0 & 0 \\ 0 & 0 & 0 }
\\ & + \frac{\gO_{R}}{2} {\bf S}_{x}\cos\p{2k_{L}\hat x} - \frac{\gO_{R}}{2} {\bf S}_{y}\sin\p{2k_{L}\hat x}
\end{split} \end{equation}
The matrices ${\mathbb I}, {\bf S}_{x,y,z}$ are the $3\times3$ identity and spin matrices in the basis $(m=1,0,-1)$,
$\hbar$ is Planck's constant divided by $2\pi$, $m$ is the atomic mass, $\delta$ is the effective detuning between states $m=0$ and $m=-1$, $\delta+\hbar\omega_q$ is the effective detuning between states $m=1$ and $m=0$, $\Omega_R/\hbar$ is the Rabi frequency of the Raman lasers, and $k_L$ is the recoil from the Raman lasers.  In the experiment $\omega_q$ is tuned via the quadratic Zeeman effect.

In the limit where $\hbar\go_{q}$ is large, the $m=1$ state is far off resonance and decouples.  The resulting energies are
\begin{widetext}
\begin{equation} \label{eq:disp}
\begin{split}
E_{\pm}\p{\vk} &= \frac{\hbar^{2} \vec{k}^{2}}{2m} \pm \sqrt{\p{2\frac{\hbar^{2} k_{L} k_{x}}{2m} + \frac{\gd}{2}}^{2} + \p{\frac{\gO}{2}}^{2}}+{\cal O}\left(\frac{\Omega_R}{\omega_q}\right),\quad
E_1\p{\vk}=\frac{\hbar^2}{2m} \p{\vk-3 k_L\hat x}^2+\frac{3}{2} \delta+\hbar \omega_q+{\cal O}\left(\frac{\Omega_R}{\omega_q}\right).
\end{split} \end{equation}
We will solely be concerned with the lowest energy band $E_{-}(k)$. Within a semiclassical treatment, where an external potential varies slowly, the low energy Hamiltonian is formally
\begin{equation}\label{eq:main}
\hat H = -\frac{\hbar^{2}}{2m}\grad^{2}
- \sqrt{\p{\frac{\gO}{2}}^{2} + \p{\frac{\gd}{2}}^{2}
 -4E_{L}\frac{\hbar^{2}}{2m}\dee{^{2}}{x^{2}} - 2i\gd\frac{\hbar^{2}k_{L}}{2m}\dee{}{x}} + V\p{x}.
\end{equation}
\end{widetext}
where $\gO = \gO_{R}/\sqrt{2}$. Throughout the rest of this work we focus on the case where $\gd = 0$.

In sections \ref{squarewell} through \ref{sec:ddw}, we study the single particle eigenstates of Eq.~(\ref{eq:main}) for a range of archetypical potentials, and compare their properties with those of standard quantum mechanics.  The results are unexpectedly rich.  We explain how to reveal the nonconventional features in experiments. Section \ref{validity} addresses the limits of validity of Eq.~(\ref{eq:main}).

There are two key properties of the dispersion in Eq.~(\ref{eq:main}).  First, the dispersion has two degenerate minima.  Typically this results in a ground state wavefunction which oscillates in space.  Second, the dispersion is anharmonic.  Some of the theorems in quantum mechanics (such as the non-existence of nodes in the ground state) are based on having a purely harmonic dispersion.  In appropriately tailored potentials, an anharmonic dispersion with a single minimum can even lead to ground-state nodes. To draw out the  role of these features, we consider the following potentials:  infinite square well, harmonic oscillator, double well.  Anharmonic dispersion are ubiquitous in lattice systems, but we are unaware of an analogous study.  Lattice systems have the additional feature that momentum space is periodic.

\section{Techniques}\label{tech}
We adimensionalize Eq.~(\ref{eq:main}), scaling all energies by $E_{L} = \hbar^{2}k_{L}^{2}/2m$ and lengths by $k_{L}^{-1}$. We restrict ourself to one dimension (1D), assuming that a tight trap has frozen out motion perpendicular to the $\hat x$ direction.  Extending the discussion to the three-dimensional case is straightforward, but the interesting results already appear in 1D.

In terms of the dimensionless variable $y=k_L x$, the dimensionless Hamiltonian becomes
\begin{equation}\label{eq:adim}
\hat {\cal H} = \br{-\dee{^{2}}{y^{2}} - \sqrt{\p{\frac{\bar \gO}{2}}^{2} -4\dee{^{2}}{y^{2}}}} + {\tilde V}\p{y}
\end{equation}
where $\bar \gO = \gO/E_{L}$, and ${\tilde V}\p{y}=V(y/k_L)/E_L$.

To numerically study Eq.~(\ref{eq:adim}), we discretize space, and write the operator $\partial_x$ as a matrix, using finite difference approximations of various orders.  We then numerically calculate the square root by standard algorithms. We verify that our results are independent of the discretization grid and the order of our approximation.

\section{Infinite Square Well}\label{squarewell}

The simplest potential to investigate is the infinite square well.  This is most easily defined by taking the limit $V_0\to\infty$ of the finite square well
\begin{equation}
{\tilde V}\p{y} = \left\{\begin{array}{lc}
0 & -k_{L}L/2 < y < k_{L}L/2
\\ V_0  & \mbox{otherwise}
\end{array}\right.
\end{equation}
It should be emphasized that even in the limit $V_0\to\infty$, the boundary condition at the edge of the well is not simply that the wavefunction vanishes, otherwise the operator in Eq.~(\ref{eq:adim}) is not self-adjoint.  If one discretizes space as described in Sec.~\ref{tech}, and maintains a finite but large $V_0$, one automatically produces a self-adjoint Hamiltonian.

\begin{figure}[ht]
\centering
   \includegraphics[width=\columnwidth] {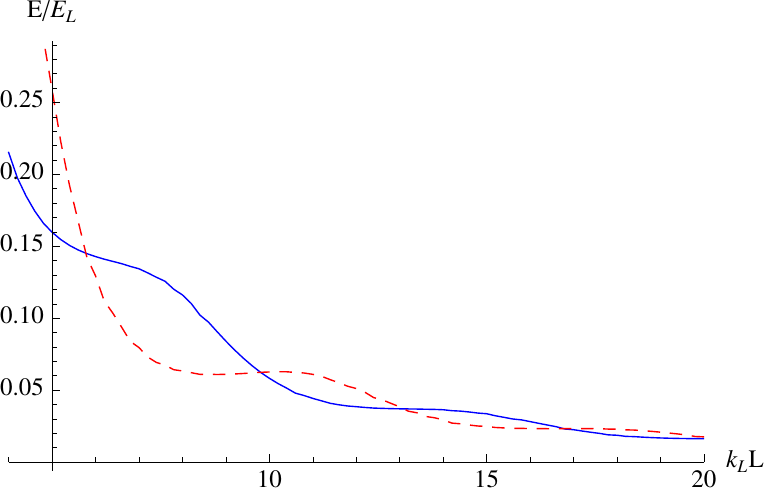}
\caption{(color online) Energies of the lowest symmetric (solid blue line) and antisymmetric (dashed red line) eigenstates of  Eq.~(\ref{eq:main}) taking $\gO = 2E_{L}$, $\delta=0$ and treating $V$ as an infinite square well of width $L$. Comparing to Fig.~\ref{fig:SquareWellNNodes} one can see the states interchange as the ground state when it becomes energetically advantageous to add another node.}
\label{fig:SquareWellEs}
\end{figure}

\begin{figure}[ht]
\centering
   \includegraphics[width=\columnwidth] {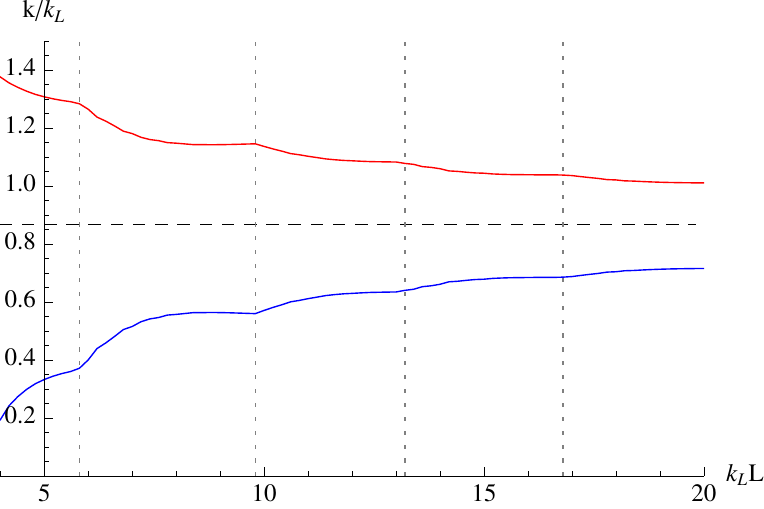}
\caption{(color online) Momenta in the ground state of the square well as a function of well size (see Fig.~\ref{fig:SquareWellEs} for parameters). The solid lines show the 
Fourier components of
the ground state while the dashed line between them is $k_{min}/k_{L}$. The dotted vertical lines denote positions where the symmetry of the ground state changes.}
\label{fig:SquareWellKs}
\end{figure}

The eigenstates can be classified by the number of nodes they possess.  In Fig.~\ref{fig:SquareWellEs}, we show the energies of the lowest two eigenstates as a function of the width of the potential well.  
As one increases the well width, the lowest symmetric and antisymmetric state take turns being ground state.  These crossings can be understood by noting that there is a preferred wave-vector in the problem.  In free space the lowest energy state has wave-vector $k_{\min}=k_{L}  \sqrt{1 - \p{\gO/4E_{L}}^{2}}$.  As one increases the size of the well, different numbers of half-waves of this wavevector fit into the well.  When an odd number fits best, the antisymmetric state has lower energy, otherwise the symmetric state wins.  In Fig.~\ref{fig:SquareWellKs} we show the two non-zero wavevectors of the ground-state inside the well.  There are exactly two wavevectors as the equation $E_-\p{k}=E$ can be manipulated to make  a quadratic equation.  This quadratic will have two real roots when $\bar \gO<4$ and $-1-\bar\gO^2/16 <E/E_L<-\bar\gO/2$.

\begin{figure}[ht]
\centering
   \includegraphics[width=\columnwidth] {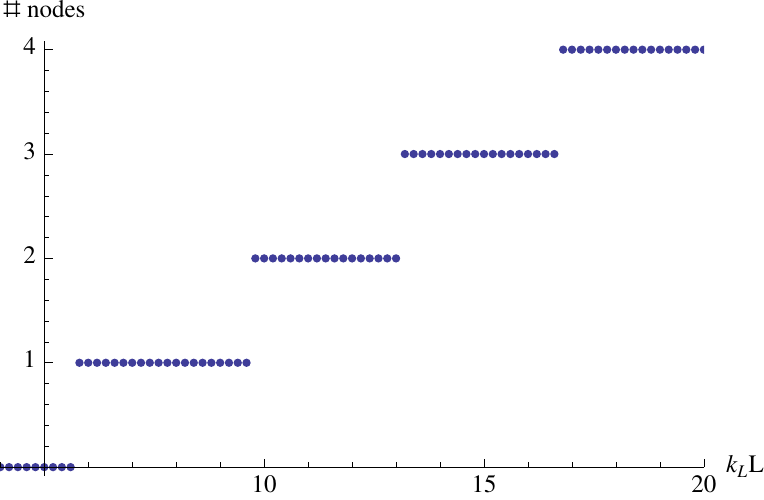}
\caption{The number of nodes in the ground state of the square well as a function of well size  (See Fig.~\ref{fig:SquareWellEs} for parameters).}
\label{fig:SquareWellNNodes}
\end{figure}

  In Fig.~\ref{fig:SquareWellNNodes} we give the number of nodes in the ground state as a function of the well width.  The contrast with usual quantum mechanics, where the ground state has no nodes, is dramatic.

\section{Harmonic Potential}
\begin{figure}[ht]
\centering
   \includegraphics[width=\columnwidth] {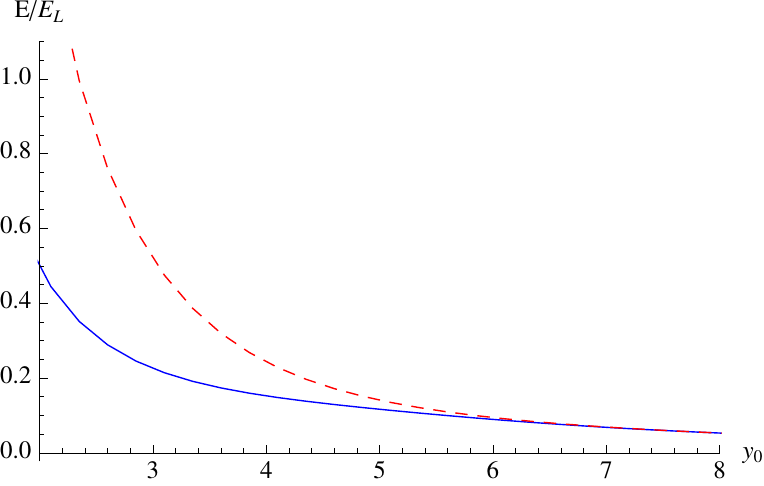}
\caption{(color online)
Energies of the lowest symmetric (solid blue line) and antisymmetric (dashed red line) eigenstates of  Eq.~(\ref{eq:main}) taking $\gO = 2E_{L}$, $\delta=0$, and taking $V=2^4 E_L (k_L x)^2/y_0^4$, corresponding to a harmonic potential with characteristic length $x_0=y_0/k_L$.}
\label{fig:SHOEs}
\end{figure}

\begin{figure}[ht]
\centering
   \includegraphics[width=\columnwidth] {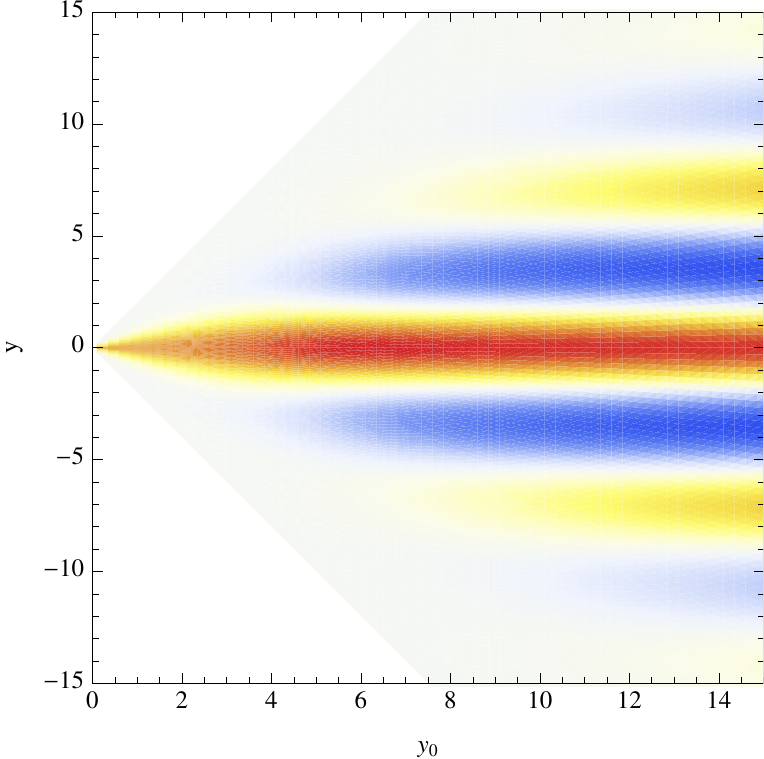}
\caption{(Color online) A density plot of the wavefunction of the ground state of the simple harmonic oscillator as a function of its width. Red and blue denote positive and negative values of $\psi$, and darker colors correspond to larger values.  Parameters are given in Fig.~\ref{fig:SHOEs}.}
\label{fig:SHOnodes}
\end{figure}

There are two reasons to study the harmonic potential.  First, the harmonic oscillator is one of the paradigmatic examples of quantum mechanics.  Second, we will be able to get further insight into the structure of the ground state by considering a canonical transformation which switches position and momentum.  Our main results, illustrated in Figs.~\ref{fig:SHOEs} and \ref{fig:SHOnodes}, are qualitatively similar to those of the infinite square well.  The main differences are: (1) The harmonic potential favors states which have a  higher amplitude at the center, resulting in a symmetric ground state.
%
%
%
(2) One cannot readily define the ``number of nodes", as the wavefunction is spread over an infinite domain, and generically oscillates an infinite number of times.  One can, however, consider analogous measures, such as how many nodes lie within a fixed number of oscillator lengths.

We consider a potential of the form
\begin{equation} \begin{split}\label{quad}
\tilde V\p{y} = \frac{y^{2}}{\p{y_{0}/2}^{4}},
\end{split} \end{equation}
and numerically calculate the eigenstates as before.

Figure~\ref{fig:SHOEs} shows the energy of the two lowest energy states as a function of $y_0$.  Unlike the square well, there are no level crossings, instead the two states simply asymptotically approach one-another.  Figure~\ref{fig:SHOnodes} shows a density density plot of the ground state wavefunction as a function of $y$ for a range of $y_0$.  As one sees, the locations of the nodes are set by the characteristic wave-vector $k_{\rm min}$, while a broader envelope is determined by the width of the potential $y_0$.

The simplest way to understand these results is to note that a Canonical transformation $y\to -i\partial_x $, $-i\partial_y \to x$ converts this to a standard quantum mechanics problem with a quadratic dispersion and a double well potential.  Thus the the Fourier transform of the wavefunction, $\tilde \psi(k)=\int dy\,e^{-ik y} \psi(y)$, is the real-space wavefunction of a conventional double well.  That is, $\tilde \psi(k)$ consists of two peaks, centered at $k_{\rm min}$ and $-k_{\rm min}$.  The width of these peaks scales as $1/y_0$.  Thus $\psi(y)=A(y) \cos(k_{\rm min} y)$, where $A(y)$ is a smooth function that falls off on a length scale of order $y_0$.

\section{Double Well}\label{sec:ddw}

\begin{figure}[ht]
\centering
   \includegraphics[width=\columnwidth] {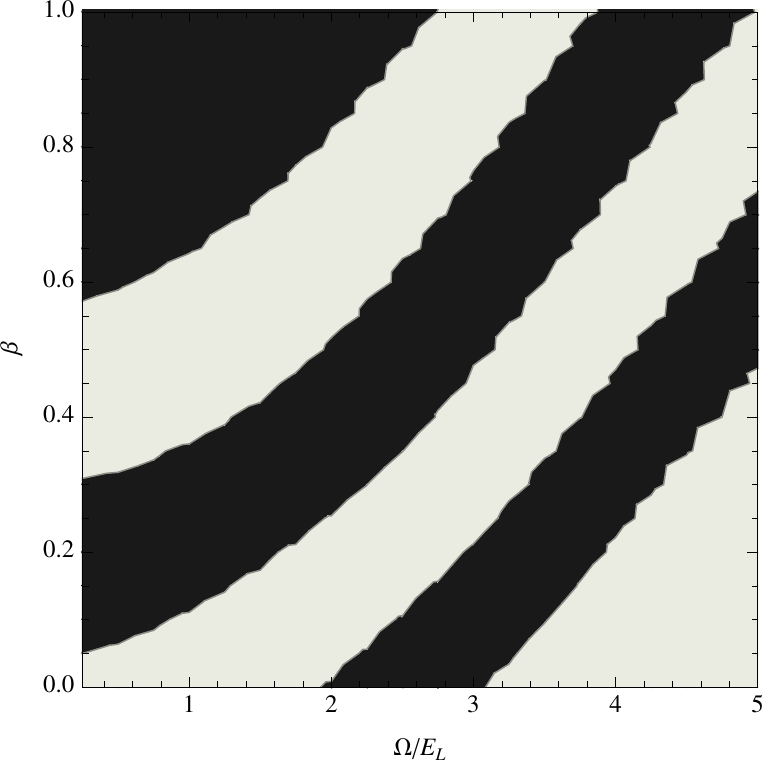}
\caption{The symmetry of the ground state of the double double well, Eq.~(\ref{double}).  Increasing the ordinate, $\beta$, increases the size of  the barrier in the real space potential, while increasing the abscissa, $\Omega/E_L$, decreases the size of the barrier in the double-well momentum space dispersion. Light areas have a symmetric ground state and dark areas antisymmetric. Here we take $y_{0} = 9, \delta=0$.}
\label{fig:doubledouble}
\end{figure}

Our final  potential is the double well, 
\begin{equation}\label{double}
V(y)= \br{\p{\frac{y}{y_{0}}}^{2}-\gb}^{2}.
\end{equation}
This is particularly interesting, as now we may have both a double well in momentum space and a double well in real space.  Either of these double-wells can take the  ground state from symmetric to antisymmetric.  Figure~\ref{fig:doubledouble} shows a cut through parameter space that illustrates this principle.

Considering only changes to the real-space potential, there are several different scenarios which lead to an antisymmetric ground state.  First, as in Sec.~\ref{squarewell}, changing the width of the real-space well changes the number of half-wavelengths of wavevector $k_{\rm min}$ that fit.   When an even number of half-wavelengths is optimal, the ground state is antisymmetric.   Equivalently, changing $k_{\rm min}$ while fixing the real-space potential, can drive a transition. Second, the bump in the double-well potential, favors wavefunctions which have a node in the center.  In regular quantum mechanics, this effect never drives the energy of the antisymmetric state below that of the symmetric state.  Here, with the non-quadratic dispersion, one can however find a level crossing.   By transforming $k\to x$ and $x\to k$ one can repeat these arguments in Fourier space.

\section{Validity of the Semiclassical approximation}\label{validity}

Our discussion so far centered on the dispersion curve Eq.~(\ref{eq:disp}) and the resulting Hamiltonian Eq.~(\ref{eq:main}). This form is achieved by applying a unitary transformation $S$ to the Hamiltonian found in Eq.~(\ref{eq:threeD}), after neglecting the decoupled off-resonance state. However, this transformation is a function of momentum, $S = S\p{k}$ and so it does not commute with the physical potential. Under this transformation, the operator in Eq.~(\ref{eq:adim}) representing the potential is 
\begin{equation} \begin{split}\label{correct}
V = S\,V_{\rm phys}\,S\dg = V_{\rm phys} + S \br{V_{\rm phys}, S\dg}
\end{split} \end{equation}
where $V_{\rm phys}=V_{\rm phys}(x) $ is the physical trap potential.  In our discussion we have neglected the second term on the right hand side of Eq.~(\ref{correct}).  Here we consider the limits of validity of this approximation.

\begin{figure}[ht]
\centering
   \includegraphics[width=\columnwidth] {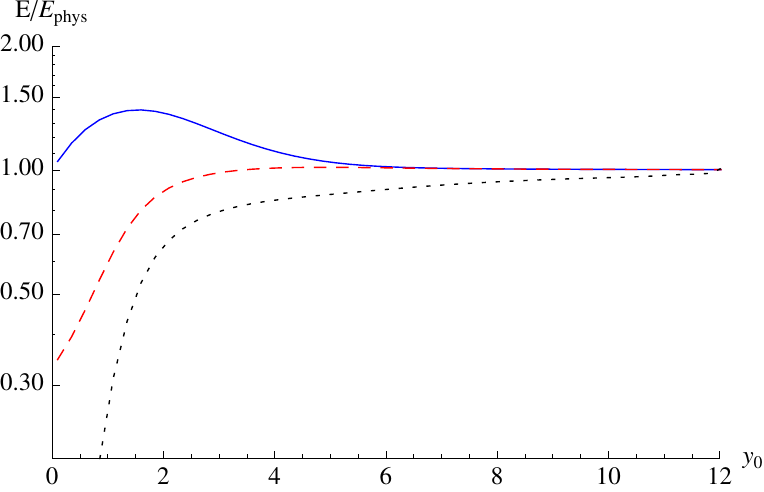}
\caption{(color online) The ratio of the energies $E$ and $E_{\rm phys}$ calculated respectively from Eq.~(\ref{eq:main}) and Eq~\ref{eq:threeD}, with $V=2^4 E_L (k_L x)^2/y_0^4, \gO=2E_{L}\omega_q=\infty,\delta=0$.  Blue solid line: ground states; Red dashed line: first excited states; Black dotted line: excitation energy.}
\label{fig:validity}
\end{figure}

As is clear from dimensional analysis, the corrections to our approximation will involve terms such as $k_L^{-1} V'(x)$.
If the characteristic scale of the changes in $V_{\rm phys}$ are large compared to $1/k_L$, then these corrections can be neglected.
 In our dimensionless units, this requires the potential $V\p{y}$ to only change on a length-scale  large compared to unity.  
 In all sections, we investigated potentials of this form, and hence we expect our results to be robust.  
 %
 Numerical investigation of the full spinor Hamiltonian, as shown in Fig.~\ref{fig:validity}, confirms that Eq.~(\ref{eq:main}) quantitatively captures the low energy physics of Eq.~(\ref{correct}).

\section{Conclusions and Outlook}

In this paper we explored the implications of the dispersion relation of Eq.~(\ref{eq:disp}) on the ground state wavefunction for several archetypical potentials. We found that the symmetry of the wavefunction could be changed by modifying the trapping potential.  This feature would also be seen for more general dispersion relations where $E(k)$ is non-quadratic and/or has multiple degenerate minima. Lin et al.'s recent realization \cite{spielman2} of the dispersion in Eq.~(\ref{eq:disp}) offers an opportunity to experimentally test our predictions.

The symmetry/antisymmetry of the ground state could be explored by either real-space probes (absorption imaging) or momentum-space probes (time-of-flight).  In particular, if $1/k_L$ is large compared to one's imaging resolution, one can simply count nodes or antinodes in the wavefunction.  In Ref.~\cite{spielman2}, $1/k_L\approx 200$nm, but this can be made longer by changing the angle between the Raman beams. 

An interesting use of time-of-flight would involve interfering outgoing waves with momentum $k$ and $-k$, giving a definitive measure of the symmetry/antisymmetry of the state.  One should be able to map out the phase diagram seen in Fig.~\ref{fig:doubledouble}: changing $\gb$ and $\Omega$ to achieve different sized bumps in the physical potential and the dispersion relation.   In particular, a clear transition should be seen going from a symmetric state localized in time and momentum for $\gO>4E_{L}$ and $\gb \ll 1$, to an antisymmetric double-well state for $\gb \sim 1$ and $\gO \lesssim 2E_{L}$.  This corresponds to a barrier whose depth is $\sim 85$ nK and a Raman coupling of order $\sim 22$ KHz.

As illustrated in Fig.~\ref{fig:SquareWellEs},  for a symmetric potential, the transitions between different symmetry ground states are true crossings, and one cannot adiabatically change from one to another.  However, if one introduces some asymmetry, these will become avoided crossings.  Depending on details, adding interactions can either further smooth out these crossings, or sharpen them, leading to further hysteresis \cite{swallow}.  Studying the role of interactions in these gases is an active area of research \cite{ho}.

\section{Acknowledgements}
This material is based upon work supported by the National Science Foundation under Grant No. PHY-1068165.

\end{document}